\documentclass{article}
\begin{document}
\begin{center}
\begin{title}
~\Large{Fermion masses and the symmetry breaking; Strong
interactions spin quantum number as the unifier of the strong and
electroweak interactions}\\
\end{title}

~

\begin{author}
~Fizuli Mamedov\\
\end{author}

~

\textit{~Institute of Physics, Baku, Azerbaijan}
\end{center}

\begin{abstract}
The origin of the fermion generations is discussed. A strong
interactions spin $I_{S}$ is introduced which unifies the quarks and
leptons as two multiplets of this spin. The electroweak vector
bosons and gluons emerge as the fused states of these fermion
multiplets.
\end{abstract}

We discussed the possibility of existence of the electroweak spin
quantum number $~I_{W} = \frac{1}{2}~$ for all fermions besides
their space - time spin  quantum number $s$ in \cite{mpr1}. This
allowed to regard the electroweak vector bosons as the fused and
condensed states of the fermions:

\begin{center}
$|s_{L}=\frac{1}{2}, ~I_{W}=\frac{1}{2}, ~m_{W}=\pm \frac{1}{2}>$
$~\times ~|s_{L}=\frac{1}{2}, ~I_{W}=\frac{1}{2}, ~m_{W}=\pm$
$\frac{1}{2}> ~=~$
\end{center}

\begin{equation}\label{fusionl}
|s_{L}= 1, ~I_{W}=1> ~+ ~|s_{L}=1, ~I_{W}= 0>
\end{equation}

\noindent Here the index $L$ indicates the left - handedness of the
fermions. The electroweak spin triplet states are represented by the
triplet of the vector gauge bosons $(\textbf{W}^{\mu})_{L}$ and the
singlet state is represented by the vector boson $(B^{\mu})_{L}$.
The fusion and condensation of the right - handed fermions leads to
the similar set of the electroweak vector bosons.

The states with $s_{L} =1$, $~(\textbf{W}^{\mu})_{L}~$ and
$~(B^{\mu})_{L}~$ couple to the left-handed fermions only and the
states with $s_{R} =1$, $~(\textbf{W}^{\mu})_{R}~$ and
$~(B^{\mu})_{R}~$ couple to the right - handed fermions only.

The $~SU(2)_{L} \times SU(2)_{R}~$ symmetry of the fermions breaks
when the $(B^{\mu})_{L}$ also couples to the right - handed
fermions. The assumption of the proportionality of the
$\textbf{W}^{\mu}$ and $B^{\mu}$ masses\footnote{We will often drop
the indices $L$ and $R$ whenever their presence is not important}
induced by the symmetry breaking to their corresponding coupling
strengths

\begin{equation}
M_{W}/M_{B}~=~g/g^{\prime}
\end{equation}

\noindent leads to the usual definition of the $Z$ boson and to the
well - known $M_{W}~-M_{Z}$ mass relation \cite{mpr1}. The fermions
gain the masses at this stage also within this model.

The three generations of the fermions could be the result of
symmetry breaking in the form of the second order phase transitions.
The most general form of the potential energy in 3 + 1 dimensions is

\begin{equation}
V(E)~=~\alpha E^{4}~+~\beta E^{3}~+~\gamma E^{2}~+~\delta E ~+~\eta
\end{equation}

\noindent where $\alpha$, $\beta$, $\gamma$, $\delta$ and $\eta$ are
the coefficients characterizing the medium. E, which is a scalar
quantity, represents the fundamental field mentioned earlier (see
\cite{mpr1}). Below the critical point of the second order phase
transitions, after the symmetry breaking, this function will have
three minima corresponding to three generations of fermions. The
further is the minimum from the minimum of the unbroken symmetry (at
$E=0$), the more prominent is its broken symmetry nature. Therefore
the fermion masses can be considered as the measure of the
brokenness of the symmetry of the original fermion states. This
possible fermion mass generation mechanism also explains why the
fermion masses are not so essential as their space - time spin and
the electroweak spin quantum numbers in the fermion - vector boson
coupling patterns.

It is natural to assume that, similar to the case with the left and
right - handed  fermions, matter and antimatter make up two
irreducible representations of the fundamental field (each with its
own left and right handed fermion representations). Therefore the
evenness of $V(E)$, $V(E) = V(-E)$ is not required.

Interestingly, this treatment of the symmetry breaking also leads to
the possibility of the existence of the second and third generations
of the superheavy electroweak vector bosons, provided that they are
the fused states of the fermions.

Could the leptons and quarks be different multiplets of another type
of spin quantum number? Indeed, if we assign an additional strong
interactions spin quantum number $I_{S}$ to the fermions, the quarks
can be regarded as the triplet $I_{S} =1$ states (quark color
states) of this spin  and the leptons will correspond to the singlet
$I_{S} = 0$ state. It is also plausible to assume that in addition
to the space - time and the electroweak spin components, fermions
also have two additional components, each of which has $I_{S}
=\frac{1}{2}$ besides other possible quantum
numbers\footnote{Closeness of the space - time and the electroweak
spin components is clear from \cite{mpr1} and will be further
supported by the discussions in this work, although they are not so
close that to have two the same quantum numbers as in the case of
the strong interactions components}. The states $|m_{S1} =\pm$
$\frac{1}{2}> \times |m_{S2} = \pm \frac{1}{2}>$ (with all other
quantum numbers the same) can be separated into the strong
interactions (color) triplet quarks and the strong interactions
singlet leptons. The fact that there are approximately three times
more quarks than leptons in nature (in the original state of the
stars) is consistent with this construction.

At the stage, when $\textbf{W}^{\mu}$ triplet are not coupled to the
fermions, the hypercharge $Y$ is the only charge of the fermions and
their sum is zero for the upper members as well as for the lower
members of the electroweak doublets \cite{mpr1}:

\begin{center}
$Y(\nu)~+~ Y(u) ~\times ~3~ =~ -1~+~\frac{1}{3} ~\times ~3~=~0$
\end{center}

\begin{equation}\label{str_int_spin_repr}
Y(e)~+~Y(d) ~ \times ~ 3~=~-1~+~\frac{1}{3}~ \times ~3~=~0
\end{equation}

\noindent In other words, ($\nu, ~u_{r}, ~u_{b}, ~u_{g}$) and ($e,
~d_{r}, ~d_{b}, ~d_{g}$) manifest themselves as two different
representations (not irreducible) of the strong interactions spin
$I_{S}$, as expected (the quark indices $r, ~b ~and ~g$ stand for
the red, blue and green colors). Naturally, this explains the
original hypercharge values of the quarks and leptons. It is
important to mention that the right - handed fermions and the left -
handed fermions have their own individual sets of these
representations. At the stage of the $~SU(2)_{L} \times SU(2)_{R}~$
symmetry breaking, the charges of the left - handed fermions gain a
$~\pm \frac{1}{2}~$($\times e$) contribution due to the coupling to
the triplet of the $\textbf{W}^{\mu}$ bosons and the charges of the
right - handed fermions gain the same contribution due to the $Y =
\pm 1$ couplings of the $(B^{\mu})_{L}$ to these fermions ('+'
contributions for the upper members of the doublets and '-'
contributions for the lower members of the doublets). With no right
- handed fermion doublets around (in terms of the hypercharges the
left - handed doublets still exist after the symmetry breaking), the
representations of the strong interactions spin now become all four
fermions of one generation together, up to the masses of the
particles:

\begin{center}
$Q(\nu) + Q(e) + Q(u) \times 3 + Q(d)\times 3 =$ $ 0 -1 +
\frac{2}{3} \times 3 -\frac{1}{3} \times 3 = 0$
\end{center}

\noindent The representations of the strong interactions spin change
after the breaking of the $~SU(2)_{L} \times SU(2)_{R}~$ symmetry,
the strong interactions also experience the consequences of the
symmetry breaking.

The lepton - lepton type fusion leads to the electroweak gauge
bosons discussed in \cite{mpr1}. The quark -  quark type fusion
leads to the following states (we omit non - essential indices in
this equation, see Eq.(\ref{fusionl})):

\begin{center}
$|m_{W}=\pm \frac{1}{2}, ~I_{S}=1> \times$
 $|m_{W}=\pm\frac{1}{2}, ~I_{S}=1>~=$
\end{center}

\begin{center}
$|I_{W} ~singlet, ~I_{S}~ singlet> ~+ ~|I_{W} ~triplet,$
$I_{S}~singlet>~ +$
\end{center}

\begin{equation}\label{vectorbosons}
~|I_{W} ~singlet, ~I_{S} ~octet> ~+~|I_{W} ~triplet, ~I_{S} ~octet>
\end{equation}

\noindent (electroweak spin singlet + electroweak spin
triplet)$\times$ (strong interactions spin singlet) states are
represented by 4 electroweak interactions vector bosons,
(electroweak spin singlet, strong interactions spin octet) states
are represented by 8 gluons. As one might expect, the strong
interactions octet states, the gluons, do not couple to the strong
interactions singlet states, to the leptons. (electroweak spin
triplet, strong interactions spin octet) states, the last term in
Eq.(\ref{str_int_spin_repr}), either do not couple to the fermions
(similar to the $(\textbf{W}^{\mu})_{R}$) or they couple very weakly
(in the latter case quite exotic particles, color changing photons
would exist).

For the electroweak spin singlet gluons the disappearance of the
right - handed doublets after the symmetry breaking is not an
obstacle and they couple with equal strength to both left - handed
fermions and right - handed fermions. Different from the $B^{\mu}$,
the strong interactions spin octet gluons also couple with the equal
strength to the upper and lower members of the fermion
doublets\footnote{The incompatibility of the tendency of the strong
interactions spin octet states to couple symmetrically to the
fermions  with the absence of the right - handed fermion -
electroweak triplet state couplings might well be an obstacle for
the coupling of the (electroweak spin triplet, strong interactions
spin octet) states to the fermions}. These equal gluon couplings,
especially the equal left - right couplings, must be the reason for
their masslessness after the breaking of the $~SU(2)_{L}\times$
$SU(2)_{R}~$ symmetry. The masslessness of the gluons could well be
connected also to the fact that the left - handed and the right -
handed fermions still constitute two unmixed representations of the
strong interactions spin even after the symmetry breaking. The
electroweak interactions bosons couple differently to the left -
handed and right - handed fermions and gain mass as a result of the
symmetry breaking.

Interestingly, all the vector bosons with $m_{W} = 0$, gluons, the
$Z$ bosons and the photons carry the interactions between the
fermions of the same type, between the particles with the same
handedness, $m_{W}$ and generation, whereas the vector bosons with
$m_{W} = \pm 1$, $\textbf{W}^{\pm}_{\mu}$ couple the upper and lower
members of the fermion doublets which have different values of
$m_{W}$. $\textbf{W}^{\pm}_{\mu}$ also induce the intergenerational
couplings between the quarks with the differing values of $m_{W}$.
Thus it is the $m_{W}=0$ values of the corresponding intermediaries
that leads to the flavor conservation in the strong interactions and
the absence of the flavor changing neutral electroweak currents.

The fermions of the different values of $m_{W}$ also have different
masses. This indicates to the existence of the connection between
the electroweak spin quantum numbers of the particles and their
masses. Also, the quarks with the larger $Y_{R}$ (the index $R$
stands for the right - handedness), $~u,c,t~$ make up heavier
sequence of masses compared to the quarks of the smaller $Y_{R}$,
$~d, s, b~$ and the leptons with $Y_{R} = 0 $, the neutrinos are
massless.

$\textbf{W}_{\mu}^{\pm}$ carry the interactions between the fermions
of the different $m_{W}$ and consequently of the different mass. It
is the capacity of the vector bosons with $m_{W}\neq 0$,
$\textbf{W}_{\mu}^{\pm}$, to couple the fermions of the different
masses that produces the intergenerational couplings of the quarks,
the Cabibbo mixing of quarks. $\textbf{W}_{\mu}^{\pm}$ do not induce
intergenerational couplings for the leptons. Perhaps, it is outside
their capacity to couple the massive fermion of one generation to
the massless fermion of another generation.

\begin{center}
\textbf{Conclusion}
\end{center}

The three fermion generations could be the reflection of the three
minima of the potential energy built out of the fundamental field,
emerging after the symmetry breaking. The introduction of the strong
interactions spin quantum number allows to consider the strong
interactions and the electroweak interactions in a unified form. The
electroweak interactions, the strong interactions and the space -
time properties of the particles are intricately connected to each
other. The separation of these attributes of the particles is always
conditional. The masses of the particles, their space - time related
quantum number are generated due to the electroweak symmetry
breaking. The leptons and quarks can be considered as two multiplets
of the strong interactions spin. The electroweak triplet of the
vector bosons, $\textbf{W}^{\mu}$ induce the intergenerational
couplings only between the strongly interacting particles, between
the quarks. The particles with $m_{W} =0$, the gluons, the $Z$
bosons and the photons do not couple fermions of the different
generations as well as of the different values of $m_{W}$.

~

\end{document}